\documentclass[aps,prb,preprint]{revtex4}
\usepackage{graphicx}

\begin{document}
\title{Experimental and theoretical study of the influence of disorder
    on diffuse first-order phase transitions: NaNbO$_3$: Gd and
    KTaO$_3$: Li as examples}

\author{M.S. Prosandeeva}

\affiliation{Institute of Theoretical and Experimental Biophysics,
Puschino, Russia}

\author{ S.I. Rayevskaya, S.A. Prosandeev, I.P. Raevski}

\affiliation{Rostov State University, Rostov on Don, Russia}

\author{S.E. Kapphan}

\affiliation{University of Osnabrueck, Osnabrueck, Germany}

\date{\today}

\begin{abstract}

We consider consequences of local disorder in systems experiencing
first order phase transitions. Such systems can be of rather
different nature. For example, manganates showing gigantic
magnetoelectric effect, doped antiferroelectrics or biomembranes.
Monte-Carlo computations performed have shown that the disorder
increases the temperature interval where the high- and
low-temperature phases coexist and this provides thermodynamics in
the disordered systems distinct from the thermodynamics of classical
homogeneous systems.

\end{abstract}

\maketitle

\section{Introduction}

The effect of disorder on macroscopic properties of crystals is a
topic of great interest because of fundamental problems, and due to
numerous new physical phenomena evidenced in complex not fully
ordered materials. For instance, PbMg$_{1 / 3}$Nb$_{2 /
3}$O$_{3}$-PbTiO$_{3}$ solid solutions show remarkable piezoelectric
properties [1], which allow transforming the mechanical energy to
electrical and back that is used in a wide spectrum of applications,
from automobile industry to device implantations in human bodies.
Other examples are perovskite-type manganates presenting giant
magnetoresistent effect. The phase transition between the dielectric
and metal phases is diffuse and contains some features (heat
capacity) of first order phase transitions and some (strong
fluctuations) of second order (see detailed discussion of this
problem in Ref. [2]). We add to this list also a biological example,
a lipid system in living cells. The main phase transition between
the solid and gel states of the lipids is diffuse and the
fluctuations are strong that reminds one of the case of manganates
[3,4].

Recently, one more example was found in a solid solution of
antiferroelectrics, NaNbO$_{3}$:Gd [5-7]. There are at least six
first order phase transitions in pure NaNbO$_3$~ but the main
permittivity maximum corresponds to an
antiferroelectric-to-antiferroelectric transition between two
orthorhombic phases, P and R (Fig. 1). This peak in the dielectric
permittivity is due to the coupling between the AFE and FE order
parameters [5]. The thermal hysteresis appears at the P-R phase
transition as straight vertical lines on the permittivity versus
temperature dependence corresponding to abrupt change of the order
parameter. Such a picture is classic. Doping NaNbO$_{3}$ with Gd
results in changing the thermal hysteresis shape: the vertical lines
on the permittivity versus temperature dependence become inclined
and there appear tails at the bottom and top of these straight
inclined lines (Fig. 1). It seems that the inclination of these
lines can be a good measure of the degree of disorder. It is
remarkable that there is a threshold Gd concentration at which the
thermal hysteresis disappears abruptly and the dielectric
permittivity maximum resembles one in relaxors [7] (Fig. 1c). In
order to show that the phases coexist at the diffuse phase
transition, a special experiment has been performed [6]. Dielectric
permittivity was recorded in the cooling run until some temperature
$T_{t}$ and, then the permittivity was measured in the heating rate.
This experiment has shown that the thermal hysteresis loop area
depends on $T_{t}$ that is consistent with the idea of phase
coexistence. If the phase transition were sharp then one would
expect the form of the temperature hysteresis loop unchanged in the
sense that the experimental points would follow the same path
independently of $T_{t}$. This experiment resembles well-known
results for the \textit{field} hysteresis of magnetization in
locally disordered magnetic systems studied by Preisah [16]. He
assumed that there was a distribution function of local fields, and,
due to this, the measured \textit{field }hysteresis loops of
permittivity were inclined and the slope of this inclination
depended on the distribution function shape. We want to apply a
similar idea in order to describe the \textit{temperature
}hysteresis loops in locally disordered systems like NaNbO$_{3}$:Gd.

The present study contains new experimental data on the systems
experiencing a diffuse first order phase transition and we will
provide a description of this diffuseness in the framework of the
Monte Carlo computation and a mean field theory. We will consider
two cases, NaNbO$_{3}$:Gd with the diffuse antifferroelectric -
antiferroelectric phase transition and KTaO$_{3}$:Li with a diffuse
first order paraelectric - ferroelectric phase transition.
KTaO$_{3}$ is a quantum paraelectric. This means that the phase
transition in it is suppressed by zero-point quantum atomic
vibrations [8]. The substitution of K by Li results in a
ferroelectric phase transiton but, due to the Li impurity disorder,
this phase transition is diffuse.

The understanding of the phenomena discussed above requires
developing a model, which takes into account the degree of disorder
and it should also describe the classic first order phase transition
in the absence of the disorder. Semenovskaya and Khachaturyan [9]
considered a two-dimensional dipole system containing defects
influencing a phase transition between the paraelectric and
ferroelectric phases. They found out that the defects help the
system avoiding the hysteretic phenomena by finding passes over
metastable states in a random network, which have only comparatively
small potential barriers between each other. Further studies of Wang
et al within the same approach [10] showed that dielectric
permittivity diffuses in such a system, and there appears a number
of relaxators, which produce a frequency dependence of dielectric
permittivity. Qian and Bursill [11] considered the smoothening of
the phase transition in PbMg$_{1 / 3}$Nb$_{2 / 3}$O$_{3}$ as a
result of random fields on the Pb sites produced by Mg and Nb.

Khomskii and Khomskii [12] performed Monte-Carlo computations in
which a correlated occupation was assumed (the probability for a
defect to join a cluster depended on the local environment of this
defect). They obtained that, under such a restriction, large
clusters appear instead of a mist of small ones. In the author's
opinion, the distribution of the percolation clusters by the size
and shape may influence the macroscopic properties of manganates.

Smolenski et al [13] considered effects of disorder in ferroelectrics. They
introduced a space distribution function of Curie temperatures, which
allowed the authors to describe the main feature of the temperature
dependence of dielectric permittivity in relaxors, the rounded dielectric
permittivity peak. The starting point in their theory was a second order
phase transition. A theory of a first order phase transition close to the
second one [2] was also considered but without the disorder.

Doniah has proposed a useful mathematical approach to the problem of
the description of the main phase transition in a lipid system [14].
He considered lipids as two-state (Ising) subsystems embedded into
an effective field, which appears due to entropy contributions
\textit{inside} these subsystems. As a result, the effective
Hamiltonian includes the field, which depends on temperature. This
effective Hamiltonian allows describing the interacting lipids as a
set of Ising subsystems in a temperature dependent field. The
effective interaction was meant to be due to interchain
interactions. Suger et al [3] and Heimburg [4] considered surface
tension as the source of the interaction among the two-state
subsystems. The disorder effects have not been addressed.

The disorder effect in NaNbO$_{3}$:Gd and KLT appears due to the randomness
of the impurity distribution in the lattice, and the randomness of the local
fields and strains [7]. In the case if one introduces a dependence of the
soft-mode vibration frequency on the space coordinate then the soft-mode
frequency at different coordinates can vanish at different temperatures.
This fact provides an opportunity to introduce a distribution function of
local Curie temperatures (the meaning of this term will be clarified below)
that has been done by Smolenski et al. for relaxors [13] and we will extend
this idea for the general case of diffuse first order phase transitions. The
use of the Curie temperature distribution function allows one to substitute
the complex problems of the real space averaging of macroscopic quantities
in locally disordered media by configuration averaging. In the present
study, we will explore this idea performing Monte-Carlo computations and
deriving a mean field theory. Our goal is developing a simple model
describing the main, most important features, of the diffuse first order
phase transitions arising due to local disorder.

\section{Experimental}

Full details of the preparation of the NaNbO$_{3}$:Gd crystals and
their characterization have been described elsewhere [5,15]. Fig. 1
shows the tendency in the dependence of the temperature hysteresis
on the Gd concentration. In a first-order phase transition, the AFE
order parameter experiences a jump and, as a result, the dielectric
function exhibits a step seen in Fig. 1a. The diffusion of this step
seen in Fig. 1b can be considered as a result of the distribution of
the Curie temperatures (similar to the distribution of local fields
in the Preisach model):

\begin{equation}
\label{eq1}
f\left( {T_c } \right) = \frac{2}{\sqrt \pi }e^{ - \left( {T_c - T_{c0} }
\right)^2 / \Gamma ^2}
\end{equation}

\noindent where $T_{0}$ is an average (macroscopic) Curie
temperature, $T_{c}$ the local Curie temperature, and $\Gamma $ the
distribution function width. The convolution of this function with
the step function yields:

\begin{equation}
\label{eq2}
\begin{array}{l}
 \varepsilon '\left( T \right) = \varepsilon _0 \left( T \right) + b\left( T
\right)\int {f\left( {T_c } \right)\theta \left( {T - T_c } \right)dT_c = }
\\
 = \varepsilon _0 + Berf\left[ {\left( {T - T_{c0} } \right) / \Gamma }
\right] \\
 \end{array}
\end{equation}
where \textit{$\varepsilon $}$_{0}(T)$ and $B(T)$ are monotonic
functions of temperature, which, in the first approximation (not far
from the Curie temperature), can be given by linear functions;
$\theta (x) = 0$ at $x < 0$ and 1 at $x > 0$; $\mathrm{erf}(x)$ is
the error function.

Fig 2 shows the fit of expression (\ref{eq2}) to the experimentally
observed \textit{$\varepsilon $'(T)} in NaNbO$_{3}$:Gd for x=0.09.
The dashed lines present the dependence that one would expect if the
phase transition were not diffuse. The fit (solid lines in Fig. 2)
shows that the width of the distribution function for x=0.09 is
about 27 K on heating and 35 K on cooling, which is nearly
comparable with the hysteresis width, 44 K. We found that the
distribution function width decreases with decreasing Gd
concentration. For pure NaNbO$_{3}$, the width of the distribution
function is negligible. Notice that the distribution function width
can be easily found from the slope of the thermal hysteresis lines:
$ctg\alpha = \Gamma / \left( {T - T_{c0} } \right)$.

We have measured also the temporal dependence of the temperature
hysteresis and found that the hysteresis width and slopes depend on
the time but the tendencies obtained remain valid even after rather
large measuring time. This point will be a subject of a separate
publication.

Our second experiment is for K$_{1-x}$Li$_x$TiO$_3$~ (KLT) single
crystals. The samples and setup have been characterized in [17]. The
integral Second Harmonic Generation (SHG) intensities were obtained
for different KLT single crystals with x=0.022, 0.036, 0.043 and
0.063 in the melt (Fig. 3). The present study treats the data
obtained with the help of the expression, which is similar to
expression (\ref{eq2}):

\begin{equation}
\label{eq3}
S\left( T \right) = S_0 (T) + B(T)erf\left( {\frac{T - T_{c0} }{\Gamma }}
\right)
\end{equation}

\noindent where $S_{0}$ and $\Gamma $ are monotonic (in our case,
linear) functions of temperature. The fit of this expression is
shown in Fig. 3. One can see that, in all the studied cases, there
is a temperature interval, at which two phases (high-temperature and
low-temperature) coexist. The distribution function width is
especially large on cooling at $x=0.063$, and it is much smaller on
heating at the same $x$.

\section{The model}

\subsubsection{Monte-Carlo computations}

We consider a two-state system in a temperature dependent field as
it was introduced by Suger et al [3] and Heimburg [4]:

\begin{equation}
\label{eq4}
H = \sum\limits_i {\Delta _i s_i^z } - \gamma \sum\limits_{ij} {\left|
{s_i^z - s_j^z } \right|}
\end{equation}

\noindent where $\Delta _i = \left( {T - T_{0i} } \right)$, $T_{0i}
$ is a local Curie temperature at which the first-order phase
transition would happen if the Ising two-state subsystems were not
interacting, $s^z$~ is a $z$-component of the spin operator
corresponding to the $i$-th site. The difference of our approach
from that used by Suger et al [3] and Heimburg [4] is the
distribution of $T_c$~ with a given Gauss type function $f(T_{0i})$.

The first term in (\ref{eq4}) describes the chemical potential of a subsystem at the
i-th site in the Ising network. The second term is the energy arising due to
the boundary between the old and new phases.

The computation within the Metropolis scheme has been done for a
two-dimensional triangular lattice, 500x500. The results are shown
in figure 4. One can see that smoothening the phase transition
results in a larger inclination of the dependence of the order
parameter on temperature. The inclination of the order parameter
dependence on temperature implies that there is a temperature
interval, in which the high- and low- temperature phases coexist.
This interval can be estimated as $\Gamma$. Figure 5 shows snapshots
obtained for $\Delta=0.1 k_B T$, $\gamma=0.4 k_B T$, $\Gamma=0.1 k_B
T$ (a), and $\Gamma=0$ (b). One can see that the cluster's size
obtained with the Gauss smoothening is larger than that without it.
In order to find an analytical criteria, in the next section, we
consider a mean field approximation of the model considered.

\subsubsection{A mean field approximation}

It is assumed in the mean field approximation that each molecule is in an
average field, which substitutes the complex two-particle interactions. The
local field can be obtained as a derivative of the energy of the system with
respect to the site occupation number, $n_{i}$, which takes one of the two
possible values, 0 and 1 (for the high-temperature and low-temperature
states, correspondingly). The total energy of the system as a function of
the occupation numbers can be written in the form

\begin{equation}
\label{eq5} H=\delta\sum{n_i} + \gamma \sum{[n_i+n_j-2n_in_j]}
\end{equation}

The interaction term here is written in the form suggested by Doniah
[11]. This term gives the same values as $\left | n_i - n_j
\right|$~ in expression (\ref{eq4}) at the values 0 and 1 for the
occupation numbers, $n_i$~ and $n_j$. The average local field can be
found now by differentiating:

\begin{equation}
\label{eq6} h=-\left< \frac{dH}{dn_i} \right> =-\Delta - \gamma \sum
\left< 1-2n_j \right> = -\Delta - 2 \gamma z \left( \frac{1}{2} - n
\right)
\end{equation}

\noindent where $z$ is the coordination number, the number of the
nearest neighbors, $n$ is the average occupation number (the order
parameter):

\begin{equation}
\label{eq7} n=\frac{1}{1+e^{-h/k_B T}}
\end{equation}

As the average field depends on $n$ [see expression (\ref{eq6})],
equation (\ref{eq7}) provides a possibility to find $n$ as a
function of temperature. Solutions of this equation are shown in
Fig. 6. At $\gamma > 2k_B T /z$, $n(T)$ does not exhibit a jump and
continuous changing the order parameter takes place. Above this
critical point, there appears an abrupt change of the order
parameter and the phase transition becomes first-order.

In order to study the solutions obtained, we computed the free
energy

\begin{equation}
\label{eq8} F=-k_B T ln Z + \left< H-H_0 \right>
\end{equation}

\noindent where $H_{0}$ = \textit{-hn}, and $Z$ is the partition
function:

\begin{equation}
\label{eq9} Z=1+ e^{h/k_BT}
\end{equation}
\begin{equation}
\label{eq10} \left< H-H_0 \right> = \gamma z n^2
\end{equation}
Fig. 7 shows the dependence of the averaged free energy on $n$ at
different values of the model parameters. One can see that
disordering results in the vanishing of the double-well behavior of
the free energy. In other words, disordering removes the average
barrier between the ground and metastable states. This is in very
good agreement with results of Semenovskaya and Khachaturyan [9] who
showed that disordering stabilizes connected chains of metastable
states having comparatively small barriers between each other.
Disordering also decreases the slope of the $n(T)$ dependence that
is in excellent agreement with experiments shown in Fig. 1 and with
the results of the Monte-Carlo computations presented in the
previous section.

\section{Summary}

The present study has shown that the two-state model supplemented
with a temperature dependent field gives reasonable explanation of
the diffusion of first order phase transitions. The general trend is
that disordering smoothens the abrupt temperature dependence of the
order parameter and this dependence becomes inclined instead of
vertical lines. The model gives a satisfactory fit to the
experimental data discussed.

We are grateful for grants 04-02-16103 (RFBR), À04-2.9-889 (Russian
Ministry of Education), and 01-0105 (INTAS).

\newpage
\section*{CAPTIONS}

Fig. 1. Temperature dependenced of the $\varepsilon'$~ for single
crystal NNG with $x=0$~ (1), $x=0.09$~ (2), $x=0.10$~ (3) and
$x=0.12$~ (4) measured at 100 kHz. Filled symbols correspond to
heating, empty symbols to cooling.

Fig. 2. The temperature hysteresis loop for NaNbO$_{3}$:Gd at x=0.09. The
symbols are experimental points, the solid line the result of the fit. The
dashed lines are guides to the eye.

Fig. 3. Temperature dependence of SHG intensity for
K$_{1-x}$Li$_x$TiO$_3$~ single crystals with different Li content.
Symbols correspond to the experimental data and solid lines show the
best fits of expression (3).

Fig. 4. The dependence of the order parameter, $n$, on $\Delta/k_B
T_{c0}$ obtained in the Monte Carlo computation: 1. $\gamma/k_B
T_{c0} = 0.5$, $\Gamma/k_B T_{c0} = 0$, 2. $\gamma/k_B T_{c0} =
0.4$, $\Gamma/k_B T_{c0} = 0$, 3. $\gamma/k_B T_{c0} = 0.5$,
$\Gamma/k_B T_{c0} = 1$, 4. $\gamma/k_B T_{c0} = 0.4$, $\Gamma/k_B
T_{c0} = 1$.

Fig. 5. The snapshots obtained in the Monte Carlo computation with
$\Gamma=0.1k_B T$ (a) and $\Gamma=0$ (b).

Fig. 6. The dependence of the order parameter on the reduced
temperature in the mean field model.

Fig. 7. The dependence of the average free energy on the order
parameter at $\gamma z / k_B T = 2.4$, $\Gamma / k_B T = 0.1$~ (1),
$\Gamma / k_B T = 1.0$~ (2), and $\Gamma / k_B T = 1.5$~ (3).

\end{document}